\documentclass[aps,amsmath,amssymb,showpacs]{revtex4}
\usepackage{amsfonts,amsthm}
\usepackage{nccmath}

\newcommand{\om}{\omega}
\newcommand{\ep}{\epsilon}
\newcommand{\al}{\alpha}

\newcommand{\bk}{\bf k}
\newcommand{\br}{\bf r}

\def\Xint#1{\mathchoice
{\XXint\displaystyle\textstyle{#1}}%
{\XXint\textstyle\scriptstyle{#1}}%
{\XXint\scriptstyle\scriptscriptstyle{#1}}%
{\XXint\scriptscriptstyle\scriptscriptstyle{#1}}%
\!\int}
\def\XXint#1#2#3{{\setbox0=\hbox{$#1{#2#3}{\int}$}
\vcenter{\hbox{$#2#3$}}\kern-.5\wd0}}

\def\dashint{\Xint-}
\begin{document}

\title{ Analytic and geometric properties of photoinduced effects in \\
        noncentrosymmetric crystals: photovoltaic current and optical rectification }

\author{Aleksandr Pishtshev}
%%\email[]{E-mail: ap@eeter.fi.tartu.ee}
\affiliation{Institute of Physics, University of Tartu, Riia 142, 51014 Tartu, Estonia}

\author{Nikolai Kristoffel}
\email[]{E-mail: kolja@fi.tartu.ee}
\affiliation{Institute of Physics, University of Tartu, Riia 142, 51014 Tartu, Estonia}

\begin{abstract}
\noindent
An original dispersion relation between the stationary coherent nonlinear optical responses
by current and polarisation is obtained.
The dispersion relation provides a new complimentary tool that can be employed
to study light-induced charge transport models and facilitate experimental data analysis.
It is shown that the origin of the coherent current and the dc-polarisation
induced in a noncentrosymmetric crystal under illumination is related to
the theory of the Berry phase and 
can be represented in terms of the renormalised geometric potentials.
This renormalisation originates from the extra phase difference acquired
by a carrier in the light field on the quantum transition between the electronic bands.
The gauge invariance of the corresponding expressions for the current and
the polarisation is demonstrated.
\end{abstract}


\maketitle

%% main text
%
%%%%%%%%%%%%%%%%%%%%%%INTRODUCTION%%%%%%%%%%%%%%%%%%%%%%
\section{Introduction}
Investigations of the interaction of an incident light with
the electronic subsystem of a noncentrosymmetric crystal have exposed a number of
time-independent effects of the second order in the amplitude of the external field.
One of these effects is the anomalous bulk photovoltaic effect (APE), which consists of
the generation of a directional macroscopic current in the closed circuit
of a spatially homogeneous semiconductor~\cite{Fridkin,Sturman1,vonBaltz1}.
Another related phenomenon is the optical rectification effect (ORE), which implies
that the crystal becomes polarised~\cite{Franken}.
Both of these effects are stationary second order responses to light
perturbation on the summarised zero frequency - the APE along the current
and the ORE along the polarisation. The mechanism of the APE consists of two channels
of different origin: a ballistic channel~\cite{Sturman1} and 
a coherent channel~\cite{Kristoffel1,vonBaltz2,Kristoffel2,Kristoffel3,Girshberg}.
The latter channel is independent of the carriers' scattering relaxation times
and is determined by the interference of quantum transition amplitudes~\cite{vonBaltz2},
which is related to the effective shift
of the carrier wave packet in real space~\cite{Sturman2}.

In the late 1980's, we ascertained for the first time a special relationship
between the current and the $dc$-polarisation. This relationship is given by 
the duality transformation~\cite{Kristoffel4A,Kristoffel4B,Kristoffel4C}
\begin{equation}\label{du1}
P(\om)\,=\, {\eta}^{-1} {j(\om)}
\end{equation}
which, before passing to the limit ${\eta}\,\rightarrow\,0$,
allows us to express the ORE polarisation ($P$) through the coherent current ($j$)
of the APE for a given frequency $\om$ of the incident light.
Here, the parameter ${\eta}\,$, which is a reciprocal averaged
relaxation time, mathematically plays the role of an intermediate variable
that would usually be taken to zero in the final calculations in nonlinear optics.
Equation~(\ref{du1}) has been obtained within the framework of the density matrix
by considering the equation of motion for the coordinate operator. 
In theoretical practice, the expression has allowed us to calculate the ORE polarisation
by using the result of the summation over the intermediate electronic
states~\cite{Kristoffel4A,Kristoffel4B,Kristoffel4C}.

In general, equation~(\ref{du1}) reflects the relationship between dissipative and
non-dissipative processes in the system. The product ${\eta}P(\om)$
can be interpreted as the unique source of coherent APE current;
in the limit ${\eta}\,\rightarrow\,0$, the product creates selection rules for
electron-hole generation induced by the absorption of photons with optical transition energies.
It thereby provides steady-state charge transport in a noncentrosymmetric semiconductor.
Because the processes of energy exchange between the light wave and the crystalline
medium depend also on the frequency of the electromagnetic field,
one can expect that the duality transformation in Eq.~(\ref{du1}) corresponds to
a wider dispersion relation that connects spectral distributions of the APE and ORE.
In the present work, we derive such a generalised relation.
As a practical tool, the relation will be especially useful for experimentalists
because it provides a simple procedure for relating the spectral representations
of the relevant polarisation and current to one another.

On the other hand, an interconnection between the coherent APE current and
geometrical features~\cite{Berry}
of the Bloch states~\cite{Zak,Resta} has not been completely realised.
Moreover, the key issue of the gauge invariance of the current remains unexplained.

The present study is a direct extension of our previous work~\cite{Kristoffel5} and
puts an emphasis on clarifying both the macroscopic and quantum-mechanical connections
between the APE and the ORE.
There are at least two main objectives for this work: \\
(i) We wish to relate the current and the polarisation from the perspective
of dispersion correspondence in addition to Eq.~(\ref{du1}). 
Such new correspondence, established with respect to $\om$, is of considerable importance
because it would allow direct expression, within the framework of experimentally measurable
frequency dependencies, of the coherent part of the APE current through the $dc$-polarisation
induced by the light wave. This link, in turn, makes it possible to experimentally represent
one effect in terms of the other.\\
(ii) In light of the application of Berry phase~\cite{Berry} effects
in various fields of physics~\cite{Bohm},
we wish to obtain a detailed understanding of the geometrical origin
of the coherent APE current and the ORE polarisation.
This is a natural step in analysing these macroscopic physical quantities because,
from a quantum mechanical point of view, they are determined by interference processes
in the space of Bloch states.
By making this connection, we recall that the general conception of the crystalline
polarisation is based on the Berry phase~\cite{Resta,Vanderbilt}.
It is also evident that when seeking insight into the geometrical features it will be important
to establish the relevant gauge invariance.
%
%%%%%%%%%%%%%%%%%%%%%%Theoretical details%%%%%%%%%%%%%%%%%%%%%%
\section{The dispersion relation between the coherent APE current
and the ORE polarisation}
Second order interaction processes between light and matter have been investigated
in a number of our works~\cite{Kristoffel3,Kristoffel4B,Kristoffel4C,Kristoffel5};
the results from these works will be used in our present theoretical analysis.
In the context of the duality of Eq.~(\ref{du1}), the functional formulation of
the coherent APE current can be written as a sum of Lorentzian functions:
\begin{equation}\label{fg1}
{\mathcal J}_{\al}(\om)=\, \frac{C}{\om^2} 
\sum_{\ep={\pm}1} \, \sum_{\beta={x,y}} I_{\beta} \sum_{{n,l}} \,
\int d^{3}{\bk}\,
			{\mathcal L}_{nl}({\bk},{\ep}{\om},{\eta}) \,
     {\Phi}_{ln}^{\al \beta}({\bk}) \,  
\end{equation}
%%%
where
\begin{equation}\label{fg10}
{\mathcal{L}}_{nl}({\bk},{\ep}{\om},{\eta}) \,=\, 
\frac{ {\eta} }
     { (E_{n}({\bk}) - E_{l}({\bk}) + {\ep}{\om})^{2} + {\eta}^{2} } \, ,
\end{equation}
\begin{equation}\label{fg1A}
{\Phi}_{ln}^{\al \beta}({\bk}) \,=\,
|p_{nl}^{\beta}({\bk})|^{2} \left[ f_{n}({\bk}) - f_{l}({\bk}) \right]
\text{Re}\{{F}_{ln}^{\al \beta}({\bk})\} \, ,
\end{equation}
\begin{equation}\label{fb1}
{F}_{ln}^{\al \beta}({\bk})\,=\, -\,
[ R_{ll}^{\al}({\bk}) - R_{nn}^{\al}({\bk}) ] \,+\,
\frac{1}{i}\,
\frac{\partial \, \ln (p_{ln}^{\beta}({\bk})) }{\partial k_{\al}} \, .
\end{equation}
Here $I_{\beta}$ is the intensity of an elliptically polarised light wave,
$C$ is a distinct constant,
$p_{nl}^{\beta}({\bk})$ with ${n{\neq}l}$ and $R_{nn}^{\al}({\bk})$
are the matrix elements of the momentum
($\hat{{\bf p}}=-i\,{\partial}/{\partial {\br}}$)
and coordinate ($i\,{\partial}/{\partial {\bf k}}$)
operators on the periodic Bloch amplitudes $u_{n\bk}({\br})$,
the indices $\al$ and $\beta$ denote the Cartesian components, and
$f_{n}({\bk}){\equiv}f(E_{n}({\bk}))$ 
are the Fermi occupation factors of the electronic states
with band energies $E_{n}({\bk})$.
The response by the coherent APE current is represented by
$j(\om) = \lim_{{\eta} \rightarrow 0 }\,{\mathcal J}_{\al}(\om)\,$;
correspondingly, the ORE polarisation is
$P(\om) = \lim_{{\eta} \rightarrow 0 }\,{\eta}^{-1}\,{\mathcal J}_{\al}(\om)\,$.

Observe that from the relation
$$
\left. \frac{1}{ x \,\pm\, i{\eta} }\,\right|_{{\eta} \rightarrow 0} \,=\,
\frac{\mathcal P}{x} \,\mp\,{i}{\pi}{\delta(x)} \, ,
$$
where ${\mathcal P}$ denotes the principal value, one can readily write
\begin{equation}\label{du3}
P(\om) \,=\, \frac{1}{ {\pi} } \lim_{{\eta} \rightarrow 0 }\,
{\int}^{+\infty}_{-\infty}\,
{dx}\,\frac{\eta}{(x-\om)^{2}+{\eta}^{2}}\,P(x) \, .
\end{equation}
By considering the ORE polarisation and the coherent APE current
as a dual pair with correspondence given by~(\ref{du1}), one finds
\begin{equation}\label{du6}
P(\om) \,=\, \frac{1}{ {\pi} } \,
{\dashint}^{+\infty}_{-\infty}\,
\frac{dx}{(x-\om)^{2}}\, j(x) \,
\end{equation}
where integration is defined in the sense of Cauchy-Hadamard (e.g.,~\cite{Martin,Boikov}).
Alternatively, there is another representation of~(\ref{du6})
which is related to a Cauchy principal-value integral by
\begin{equation}\label{du7}
P(\om) \,=\, \frac{1}{ {\pi} } \, \frac{d}{d\om} \,
{\mathcal P}{\!\!}{\int}^{+\infty}_{-\infty}\,
\frac{dx}{x-\om}\, j(x) \, .
\end{equation}
If one defines the Hilbert transform of a real function of the real variable
$j(\om)$ as (e.g.,~\cite{Reyna,Papoulis})
\begin{equation}\label{ht1}
{\mathcal H}\{ j(\om) \}
 \,=\, \frac{1}{ {\pi} } \, 
{\mathcal P}{\!\!}{\int}^{+\infty}_{-\infty}\,
\frac{dx}{x-\om}\, j(x) \, ,
\end{equation}
it is seen from Eq.~(\ref{du7}) that the ORE polarisation is
a derivative with respect to $\om$ of the Hilbert transform of the coherent APE current.
By using the property 
${d} {\mathcal H}\{ j(\om) \} {/}{d\om} \,=\, {\mathcal H}\{ {d} j(\om){/}{d\om} \}$,
Eq.~(\ref{du7}) can be rewritten as
\begin{equation}\label{du7A}
P(\om) \,=\, \frac{1}{ {\pi} } \, 
{\mathcal P}{\!\!}{\int}^{+\infty}_{-\infty}\,
\frac{dx}{x-\om}\, \frac{dj(x)}{dx} \, .
\end{equation}
With the help of Eqs.~(\ref{ht1})~and~(\ref{du7A}) the ORE polarisation
can be regarded as the convolution of the frequency derivative
of the coherent APE current with a power function of the form $1/{(x-\om)}$.

The above results (Eqs.~(\ref{du6}),~(\ref{du7}),~and~(\ref{du7A})) constitute
our desired dispersion relation, which connects
the frequency dependencies of the coherent APE current and the ORE polarisation.
Therefore, once we know a spectral representation of one quantity the above equations
give the relevant representation of the other.
In fact, this provides a new complementary tool for the study of  
steady-state nonlinear optical effects. 
A more practical motivation for these studies lies in their technological importance. 
For example, the phenomenon of optical rectification is of considerable interest
because it is related to the ability to generate terahertz radiation (e.g.,~\cite{Tonouchi}).

Because $P(\om)=P(-\,\om)\,$, relation~(\ref{du6}) can be also written as
\begin{equation}\label{du8}
P(\om) \,=\, \frac{1}{ {\pi} }
{\dashint}^{+\infty}_{-\infty} {dx}\,
\frac{x^{2}+{\om}^{2}}{(x^{2}-{\om}^{2})^{2}}\, j(x) \\
\,=\,
\frac{1}{ {\pi} } \, \frac{d}{d\om} \,
{\mathcal P}{\!\!}{\int}^{+\infty}_{-\infty} dx \,
\frac{\om}{{x}^{2}-{\om}^{2}}\, j(x) \,.
\end{equation}
As a simple illustration of Eq.~(\ref{du8}), the spectral frequency profile of $P(\om)$
for a ${\delta}$-shaped coherent APE current
$j(\om)=j_{0}\,[{\delta}(\om-\om_{0})+{\delta}(\om+\om_{0})]$
is given (for ${\om}\,{\neq}\,{\om}_{0}$) by
\begin{equation}\label{du9}
P(\om) \,=\, \frac{2j_{0}}{ {\pi} }\,
\frac{\om_{0}^{2}+{\om}^{2}}{(\om_{0}^{2}-{\om}^{2})^{2}} \,.
\end{equation}
%
%%%%%%%%%%%%%%%%%%%%%%The geometric properties%%%%%%%%%%%%%%%%%%%%%%
\section{The geometric properties}
%%%
Let ${\Theta}_{ln}^{\beta}({\bk})$ be the phase of the interband matrix element
of the momentum operator: $p_{ln}^{\beta}({\bk})=
|p_{ln}^{\beta}({\bk})|\,e^{i\,{\Theta}_{ln}^{\beta}({\bk})}\,$.
The first derivative of the amplitude 
${\partial |p_{ln}^{\beta}({\bk})| }/{\partial k_{\al}}$
does not contribute to the integral in Eq.~(\ref{fg1})
%%%as it is an odd function of  ${\bk}$%%%
and,
correspondingly, Eq.~(\ref{fb1}) can be written as
\begin{equation}\label{fb2}
{F}_{ln}^{\al \beta}({\bk})\,=\, -\,
[ R_{ll}^{\al}({\bk}) - R_{nn}^{\al}({\bk}) ] \,+\,
\frac{\partial {\Theta}_{ln}^{\beta}({\bk}) }{\partial k_{\al}} \, .
\end{equation}
The first term in the square brackets of Eq.~(\ref{fb2}) describes the difference
in the positions of the electron charge
for the initial and final states. 
By using the definition of a gauge potential (a Berry connection) in the ${\bk}$-space 
for the band $n$ as ${\bf{A}}_{n}{\equiv}\,{\bf{R}}_{nn}\,$, one has
\begin{equation}\label{fb3}
{\bf{F}}_{ln}^{\beta} \,=\,
{\bf{A}}_{n}\,-\,{\bf{A}}_{l}\,+\,{\rm grad}{\Theta}_{ln}^{\beta} \, .
\end{equation}

As seen from Eqs.~(\ref{fg1})~-~(\ref{fb1}) and~(\ref{fb3}), 
the expression for the current in the geometrical sense is formulated
in terms of vector potentials and not fields, and so the relevant gauge invariance
must be verified. 
Clearly, the current is an observable physical quantity and should be
independent of the freedom to change the phase of the Bloch wave function
arbitrarily at each point in $\bk$-space.

The gauge symmetry of the current is simply stated from Eq.~(\ref{fb1})
by considering the quantity ${\bf{F}}_{ln}^{\beta}$
to be invariant under a local gauge transformation:
$u_{n\bk}({\br}) \longrightarrow\,e^{i\,\varphi_{n}(\bk)} \, u_{n\bk}({\br})$
where the quantity $\varphi_{n}(\bk)$ is an arbitrary phase of the Bloch amplitude.
According to the sense of the Berry potential, we can call such a gauge geometric.
As a direct consequence, we reveal the physical role played by the gradient term 
in Eqs.~(\ref{fb1}) and~(\ref{fb3}) as follows:
the term maintains the geometric gauge invariance of the coherent APE current
and the ORE polarisation,
and hence eliminates the arbitrary  non-gauge contributions
caused by phases of the Bloch amplitudes. 
In view of Eq.~(\ref{fb3}), this allows us to conceptually treat
the phases of the matrix elements of the momentum operator as compensating fields.

Now that we have described the roles of both components in Eq~(\ref{fb3}),
it is possible to gain better insight into the basic properties of 
the quantity ${\bf{F}}_{ln}^{\beta}$.
First, by defining ${\Xi}_{ln}^{\beta}$ as half of ${\Theta}_{ln}^{\beta}$,
we can rewrite Eq~(\ref{fb3}) as
\begin{equation}\label{fb3A}
{\bf{F}}_{ln}^{\beta} \,=\,
{\bf{A}}_{n}^{r}\,-\,{\bf{A}}_{l}^{r}\, \, 
\end{equation}
where 
${\bf{A}}_{n}^{r} = {\bf{A}}_{n}\,+\,{\rm grad}\,{\Xi}_{ln}^{\beta}$
and
${\bf{A}}_{l}^{r} = {\bf{A}}_{l}\,+\,{\rm grad}\,{\Xi}_{nl}^{\beta}\,$,
respectively.
Furthermore, we call attention to the point that, due to the Stokes theorem,
the calculation of the difference between
the Berry phases ${\Phi}_{n}$ and ${\Phi}_{l}$, which are acquired by the wave functions of
the Bloch initial ${\psi}_{n\bk}(\br)$ and final ${\psi}_{l\bk}(\br)$ states,
does not involve the gradient terms
\begin{equation}\label{fb4}
{\Phi}_{n} \,-\, {\Phi}_{l} \,=\,
\oint d{\bf l}\,{\cdot} \left[ \,{\bf{A}}_{n}\,-\,{\bf{A}}_{l}\, \right] 
\,=\, 
\oint d{\bf l}\,{\cdot} {\bf F}_{ln}^{\beta} 
\,=\,
\oint d{\bf l}\,{\cdot} \left[ \,{\bf{A}}_{n}^{r}\,-\,{\bf{A}}_{l}^{r} \, \right] 
 \, .
\end{equation}
A key observation that directly follows from Eq.~(\ref{fb4}) is that there is 
a simple way to rewrite the quantities ${\bf A}_{n}^{r}$ by employing
the definition of the Berry connection 
\begin{equation}\label{fb5}
{\bf A}_{n}^{r}
\,=\,
\left\langle \, u_{n\bk}^{r}({\br}) \mid 
i\,{\partial}/{\partial {\bf k}} 
\mid u_{n\bk}^{r}({\br}) \, \right\rangle
\end{equation}
where the matrix elements are calculated on the Bloch amplitudes renormalised
by multiplying by the phase factor:
$u_{n\bk}^{r}({\br}) =\, u_{n\bk}({\br})\exp(-i\,{\Xi}_{ln}^{\beta})\,$.

Such a representation of the phase-renormalised Bloch states,
in which the relevant gradient terms move into the phases
of the Bloch amplitudes, leads to a new understanding of the role played by
the phases of the interband matrix elements of the momentum operator and,
therefore, allows us to draw several conclusions:\\ 
(i) From Eqs.~(\ref{fb3A})~and~(\ref{fb5}), one sees the geometrical meaning of 
${\bf F}_{ln}^{\beta}$ as the quantity representing a difference of the renormalised
potentials of the $n$ and $l$ electronic bands.\\
(ii) The phase factor $\exp(-i\,{\Xi}_{ln}^{\beta})$ provides a direct argument
that an extra nontriviality appears in the geometry of the Bloch states
of the perturbed system, a nontriviality that is of geometric phase origin
(a similar extra phase, often called "statistical", takes place in studies
of the statistics of quasiparticles entering the quantum Hall effect~\cite{A-W}).
Geometrically, the allocated phase factor is what incorporates the effect
of the other band in the Bloch amplitude of the given band 
through the half-phase of the momentum matrix element between these bands.
This, in turn, implies that
it is also possible to consider the quantity ${\Xi}_{ln}^{\beta}$ in the sense of
phenomena such as the Aharonov-Bohm effect~\cite{A-B}
as the phase difference adiabatically acquired by the electron on the transition
between the $n$ and $l$ bands.\\
(iii) These phase differences cannot be ignored because they are employed as
compensating fields to ensure that equations for the coherent APE current
and the ORE polarisation are independent of the choice of the Bloch amplitude phase.
Thus, the geometric gauge invariance of both physical quantities is provided
by a nontrivial superposition of the difference between the "bare" Berry connections
and the interband phase difference acquired in the electron transitions.

We must emphasise that the coherent photoinduced current of the APE and
the $dc$-polarisation of the ORE are completely defined by 
the gauge invariant contributions of the renormalised Berry connections
in the space of the Bloch states' quasimomenta.
Because they are determined by the geometry of this space and not by the particular
dynamics of the Bloch electrons, neither effect naturally depends on the specific scattering
mechanisms of the carriers. For the present case, the characteristic geometric features
are associated with the interference of quantum transitions between the intermediate
electronic states.
Therefore, using the terminology of the anomalous Hall effect~\cite{Sinitsyn},
such contributions can be called intrinsic contributions.

In this situation, the meaning of the relaxation parameter ${\eta}$,
which, like the Berry phase, is strictly related to the adiabatic evolution of the system,
can also be deepened. One may say that the parameter limits the time to collect
the necessary Berry phase change during the light period.
%%%
\section{Discussion}
The final question on which our study can shed light concerns the interrelation
of the analytical and geometrical properties of the coherent APE current and
the ORE polarisation. Following a causal formulation of the coherent nonlinear
photoinduced effects~\cite{Kristoffel5},
consider the evolution of the electronic subsystem
of a crystal in the external light field.
According to the concept of time ordering in the dynamic quantum multi-electron
system~\cite{Godunov,Godunov1,Godunov2} and by using the results of Ref.~\cite{Kristoffel5},
an expression giving the stationary second-order response for
an arbitrary operator $\hat{D}$
can be rewritten in terms of the Heaviside step functions ${\Theta}(t)\,$ as follows:
\begin{equation}\label{to1}
{\delta}^{(2)} \! < \!{\hat{D}}\! > \,=\, 
{i}^{-2} \, {e}^{-{\eta}t} \, 
{\int}^{+\infty}_{-\infty}\,dt_{1}\,
{\int}^{+\infty}_{-\infty}\,dt_{2}\, {e}^{{\eta}t_{2}} \\
\left.
{\Theta}(t\,-\,t_{1})\,{\Theta}(t_{1}\,-\,t_{2}) 
\left\langle
[[\tilde{D}(t),\,\tilde{V}(t_{1})],\,\tilde{V}(t_{2})]
\right\rangle 
\right|_{0} \, .
\end{equation}
Causal-like sequencing of interactions (time ordering)
in the time evolution of the system is imposed here in an explicit form
through the products of the Heaviside step functions.
In Eq.~(\ref{to1}), $V(t)$ is the perturbation caused by illumination,
the tilde means the interaction representation
and index "$0$" selects the stationary part of
${\delta}^{(2)} \! <\!{\hat{D}}\!> \,$.

A valuable property of a representation like Eq.~(\ref{to1})
is the possibility to employ a time sharing scheme~\cite{Godunov1,Godunov2}
in which the time ordering is separated into two terms - the uncorrelated
($T_{unc}$) and correlated ($T_{corr}$) parts - by using the following representation
of the function ${\Theta}(t_{1}\,-\,t_{2})\,$:
\begin{equation}\label{to2}
%%\begin{aligned}
{\Theta}(t_{1}\,-\,t_{2})  =  \frac{1}{2}\,  \Bigl ( \,
\underbrace{\,\,1\,\,}_{\quad T_{unc} } \,+\,
\underbrace{\rm{sign}(t_{1}\,-\,t_{2})}_{\quad T_{corr} } \,  \Bigr ) \, .
%%\end{aligned}
\end{equation}
Clearly, the two contributions tend to complement each other.
The first contribution, $T_{unc}$, corresponds to an independent time
approximation where the $\tilde{V}(t_{j})$ interactions are disconnected
among themselves in time. The second contribution, $T_{corr}$,
connects the $\tilde{V}(t_{j})$ at different times~\cite{Godunov1}.
By employing the Fourier transform of ${\Theta}(t_{1}\,-\,t_{2})$,
the given scheme~(\ref{to2}) can be supplemented with two 
parallelisms~\cite{Godunov,Godunov2}
\begin{align}
T_{unc} \longrightarrow\, & \lim_{{\eta} \rightarrow 0 }\,
\frac{\eta}{(E_{0}-E\,{\pm}\,{\om})^{2}+{\eta}^{2}}   
 = {\pi}\,\delta(E_{0}-E\,{\pm}\,{\om})
\, , \label{to3} \\
T_{corr} \longrightarrow\, & \lim_{{\eta} \rightarrow 0 }\,
\frac{E_{0}-E\,{\pm}\,{\om}}{(E_{0}-E\,{\pm}\,{\om})^{2}+{\eta}^{2}}
=\frac{\mathcal P}{E_{0}-E\,{\pm}\,{\om}} \label{to4}
\end{align}
where $E$ is an excited state in the electronic subsystem.

In order to see whether the coherent APE charge transport and the ORE polarisation
can be categorised according to their time-ordering properties, we shall consider
separately each contribution prescribed by Eqs.~(\ref{to2}) -~(\ref{to4}).
As a first case, we discuss the situation driven by $T_{unc}$ (i.e.,
the time-independent part of ${\Theta}(t)$) in which the sequence of the 
interactions $\tilde{V}(t_{j})$ is unimportant. 
As mentioned in Ref.~\cite{Godunov1}, the second order process can be represented
in this approximation by two independent one-step processes, which correspond in
Eq.~(\ref{to3}) to the delta function. At the same time,
as seen from Eq.~(\ref{fg10}) at ${\eta}\,\rightarrow\,0\,$, the one-photon interband
absorption processes are responsible for the generation of free carriers. 
Because they are dependent on the relevant amplitudes, these processes maintain no phase
information inside the area occupied by electrons.
This implies that we need to invoke the system's other degrees of freedom
in order to discover the asymmetry responsible for 
the rise of the coherent APE current  $j(\om)$.
In quantum mechanical systems, one available degree of freedom belongs to
the special geometric properties in the ${\bk}$-space of the Bloch states~\cite{Zak}.
As we have seen above, the microscopic origin of a macroscopic coherent current
is related to an interplay of the Berry geometric potentials and the interband
phase differences. In a medium without a centre of inversion, the sum of these quantities
provides gauge-invariant polar distributions in individual electron transitions
between the Bloch states in momentum space.

Therefore, one can conclude that the light-induced stationary coherent APE charge transport,
being of geometric origin, does not need any special time correlations between electrons.
This result is not surprising because one should expect that time-correlating processes
cannot be involved in the case under consideration due to the absence of the time-reversal
invariance associated with the bulk macroscopic current. Moreover, from the quantum mechanical
point of view correlation in time, which is initially an attribute of the incident photons
due to Bose statistics, is destroyed along with the photons themselves, which are absorbed
in order to generate the electron transitions.

A diametrically opposite situation takes place in the second case
when the time-dependent part of ${\Theta}(t)$ in Eq.~(\ref{to2}) is considered.
This case includes the time-correlating processes in the time evolution of the electron
subsystem excited by the external electromagnetic field. In a mathematical context,
the case is given by the principal value contribution in Eq.~(\ref{to4})
and driven by $T_{corr}$.
From Eqs.~(\ref{fg1}),~(\ref{fg10}),~(\ref{du1}), and~(\ref{to4}), one sees that
the light-induced ORE polarisation $P(\om)$ can be associated with
the effect of the time correlation term, $T_{corr}$.
This result implies that the interactions of a light wave with the electron subsystem
at various times must interfere to polarise a nonlinear medium. The need for such interference
can be described in a more pictorial way. The external perturbations affecting the electron
subsystem lead to local changes of the electron density distribution that, in turn, create
imbalances with respect to the corresponding density of the positive background. In the case of
the time correlations, these imbalances will be accumulated in a sequential and dependent manner
in order to involve a sufficiently large number of valence electrons in virtual transitions.
The induced polarisation is, thus, a result of a new macroscopic redistribution of the electronic
charge.

Following from the above discussion, time ordering is the result of a nonlinear effect
due to the specific correlations caused by the interactions of a light wave with
the electronic subsystem. This leads to interesting consequences with respect to
the two physical quantities related by the duality transformation in Eq.~(\ref{du1}):
one turns out to be connected with the time correlation while the other is not
(at least in the second order approximation). Simultaneously, both of them are caused
by the geometric features in the ${\bk}$-space of the Bloch states.
In addition to the explanation given in the introduction, this observation
allows us to treat the factor ${\eta}$ as a driving parameter, which shifts
the time correlated and uncorrelated contributions amongst themselves at 
${\eta}\rightarrow0$. Next, analogously to the above treatment of~(\ref{du1}),
the physical meaning of the dispersion relation in Eq.~(\ref{du6}) can also be completed.
One can say that this relation interconnects the time correlated contributions
with the uncorrelated contributions with respect to $\om$.
Obviously, such an add-on to the explanation of~(\ref{du6}) should have greater
generality because, for example, a similar treatment exists for a much different situation
(i.e., fast collisions of atoms with matter and light) ~\cite{Godunov1}.

In summary, we focused our attention on the analytical properties and the geometric meaning
of the second-order photoinduced effects in noncentrosymmetric crystals. We established an
original dispersion relation, which connects the spectral representations of the coherent APE
current and the ORE polarisation. It was shown that both phenomena are encoded by the renormalised
Berry connections (potentials). The renormalisation of the Berry connections is determined by
the interband phase differences in the quantum transitions between the Bloch states and provides
the underlying gauge symmetry in the ${\bk}$-space. We compared properties of the coherent APE
current and the ORE polarisation with respect to a physical picture in which the electron
subsystem is considered to be correlated in two different ways. It was determined that one way
takes into account the time-correlating processes created by the light wave while the other way
includes the correlations related to intrinsic geometrical features of the Bloch states.
As a result, we observed that the ORE polarisation appears to be the combined result of both ways.
Conversely, the coherent APE current is solely associated with the second way.
\section*{Acknowledgments}
The work was supported by the Estonian Science Foundation grants No. 7296 and No. 6918.
%
%%%%%%%%%%%%%%%%%%%%%%
%
%~~~~~~~~~~~~~~~~~~~  REFERENCES  ~~~~~~~~~~~~~~~~~~~~~~~~~~~~~~~~~~~~~~~~~~%
%

%%%%%%%%%%%%%%%

\end{document}